\documentclass[10pt,twocolumn]{article}
\usepackage{amsmath,amssymb,amsthm,mathtools}
\usepackage[T1]{fontenc}
\usepackage{lmodern}
\usepackage[hidelinks]{hyperref}
\hypersetup{
  pdftitle={Full Provenance on the COSS Causal Axis: An Event-Level Closure of Causal Attribution and Confused-Deputy Exclusion},
  pdfauthor={Nicola Gallo}
}
\usepackage{microtype}
\usepackage{booktabs}
\usepackage{array}
\usepackage[letterpaper,margin=0.72in]{geometry}
\usepackage{stmaryrd}
\usepackage{enumitem}
\usepackage{url}
\usepackage{xspace}

\newcommand{\Model}{\textnormal{\textsc{COSS}}\xspace}

\newcommand{\PoR}{\mathrm{PoR}}

\newcommand{\sem}[1]{\llbracket #1 \rrbracket}
\newcommand{\powerset}{\mathcal{P}}
\newcommand{\state}{\Sigma}
\newcommand{\coord}{\sigma}
\newcommand{\st}{\operatorname{st}}
\newcommand{\FP}{\mathrm{FP}}

\newcommand{\CDAF}{\mathrm{CDAF\text{-}E}}
\newcommand{\Influence}{\mathrel{\rightsquigarrow_{\!\FP}}}
\newcommand{\CausedBy}{\operatorname{CausedBy}}
\newcommand{\Realize}{\operatorname{realize}}
\newcommand{\Exec}{\operatorname{exec}}
\newcommand{\IndAuth}{\operatorname{IndependentAuthority}}
\newcommand{\Accepted}{\operatorname{Accepted}}
\newcommand{\hb}{\prec_{c}}
\newcommand{\Hist}{\operatorname{Hist}}
\newcommand{\Beh}{\operatorname{beh}}
\newcommand{\Ord}{\operatorname{Ord}}
\newcommand{\OrdSeg}{\operatorname{OrdSeg}}
\newcommand{\AuthSource}{\operatorname{AuthSource}}
\newcommand{\CDApic}{\mathrm{CDA}_{\mathrm{PIC}}}
\newcommand{\Pred}{\operatorname{Pred}}

\newtheorem{definition}{Definition}
\newtheorem{theorem}{Theorem}
\newtheorem{proposition}{Proposition}
\newtheorem{lemma}{Lemma}
\newtheorem{corollary}{Corollary}
\newtheorem{remark}{Remark}

\title{Full Provenance on the COSS Causal Axis:\\
An Event-Level Closure of Causal Attribution and\\
Confused-Deputy Exclusion}
\author{Nicola Gallo}
\date{2 October 2026}

\begin{document}

\maketitle

\begin{abstract}
This paper follows the Provenance Identity Continuity (PIC) model and its
relational extension beyond proof of possession.  The earlier formulation
represents an authority-use event as $(o,r,\ell)\in O\times R\times L$: an
operation/resource behavior situated in a causal lineage.  We make the causal
occurrence itself primitive.  Let $\mathcal E$ be the underlying occurrence
universe and let $(E,\to_c)$, with $E\subseteq\mathcal E$, be the represented
security-relevant causal graph.  Each occurrence has a unique causal history
$\Hist(e)$, so the consistent event-lineage space
$\widehat E=\{(e,\ell)\in E\times L\mid \ell=\Hist(e)\}$ is canonically
isomorphic to $E$.  The modeled execution has one causal partial order, not one
clock or one order per lineage.  Distinct lineages are distinct causal
histories inside that order and may share the same root while diverging into
incomparable branches.  PIC's $O\times R\times L$ is then the
authority-behavior projection of this causal occurrence space.  COSS
generalizes the same construction by replacing the authority behavior
$O\times R$ with an arbitrary behavior product $\mathcal B$ while retaining
that one causal execution.

On this event semantics we prove an architecture-independent confused-deputy
exclusion theorem.  Ordinary PIC continuation is required on every ordinary represented
causal edge: $C(e')\subseteq C(e)$.  Consequently every descendant connected
to an ancestor by an ordinary authority segment is bounded by that ancestor.
At an ordinary multi-source join this is
equivalent to $C(e)\subseteq\bigcap_{p\in\Pred(e)}C(p)$; authority available
on only one incoming branch therefore cannot be silently resurrected at the
join, even when all branches share a common root.  Common ancestry is weaker
than continuation identity.  Any legitimate non-conservative combination must
be an explicit authorized composition or relaxation event.  With
protected-effect mediation, an accepted effect caused by request occurrence
$q$ and connected to it by an ordinary authority segment can exercise only
authority in $C(q)$.  Thus a confused-deputy state, which requires the same request-caused effect to
exercise authority absent from $C(q)$ but independently held by the deputy, is
contradictory.  The result is universal over causally complete event models subject to the
ordinary-segment premise: for any system whose protected request-to-effect causes are faithfully
represented in the event DAG and whose ordinary authority transport satisfies
PIC, no PIC lineage-form confused-deputy effect is admissible through
ordinary continuation.

We then revisit Rajani, Garg, and Rezk (RGR).  Their capability semantics blocks
explicit designation but not all value, control-flow, or initial-heap attacks;
their full provenance semantics tracks explicit and implicit influence and
proves CDA-freedom with endorsement for their whole region calculus.  From the
causal PIC view, these attack classes are different witnesses of causal
ancestry.  Full provenance is therefore a concrete evidence mechanism for
recovering data/control edges of the common causal graph, not the foundation of
the PIC impossibility theorem.  The paper finally lifts the construction to the
COSS product of heterogeneous security-state coordinates.  ``Total'' here is
model-relative and ordinary-segment-relative: it covers all represented PIC
lineage-form CDA instances whose request/effect pair satisfies the ordinary
continuation premise, not physical channels or non-conservative transitions
outside that scope.
\end{abstract}

\section{Introduction}

This work continues the PIC sequence: the original Proof-of-Continuity model
binds authority to causal execution lineage, while \emph{Beyond Proof of
Possession} separates possession from admissible authority and makes explicit
the role of request-caused effects \cite{gallo_poc2026,gallo_relational2026}.
The present paper asks what changes when the execution occurrence itself,
rather than an operation/resource tuple, is taken as the primitive object.

The answer is a causal generalization.  A machine run consists of occurrences
ordered by causal dependence.  Operation, resource, principal, value flow,
control decision, geographic location, software provenance, and other security
facts are observations of those occurrences.  They need not define separate
times.  PIC's event $(o,r,\ell)$ is therefore read as the authority projection
of a richer causal occurrence.  COSS keeps the same causal axis and permits a
behavior $b\in\mathcal B$ and a product security state to be attached to the
same occurrence.

This viewpoint changes the confused-deputy problem.  The relevant structure is
not the wall-clock distance between a request and a privileged effect.  It is
the branch-sensitive causal structure that reaches the effect.  A request may
influence an effect through a direct call, a value, a branch, mutable state, a
queue, a database, a timer, a callback, a message, or an arbitrarily long
chain.  An authority-bearing branch may share an old root with the request yet
still be a different causal continuation after a fork; it may also be
concurrent with the request rather than ordered before it.  In the event DAG
these histories remain distinct until an explicit join.  PIC then supplies the
authority invariant: ordinary continuation cannot silently enlarge, restore,
or union the authority carried from any causal parent merely because the
parents have a common ancestor or meet in one process.

This observation yields the paper's main theorem.  If an accepted effect $e$
is a causal descendant of a request $q$, and the segment from $q$ to $e$ is
ordinary PIC continuation, then $C(e)\subseteq C(q)$.  Since an accepted effect
realizing $(o,r)$ must have $(o,r)\in C(e)$, it follows that $(o,r)\in C(q)$.
The lineage-form confused deputy requires the opposite.  Hence no such state is
admissible.  The theorem is independent of whether the intermediate cause was
called explicit designation, implicit designation, value influence, control
flow, heap influence, or asynchronous state transfer.

The word \emph{universal} is used here in a precise structural sense.  The
theorem quantifies over any system that admits a faithful, causally complete
representation of its protected execution and enforces PIC on ordinary causal
continuation.  It is not a claim that a formal model automatically observes an
unmodeled physical or covert cause.  Complete causal representation and effect
mediation are realization obligations.

After establishing this result, we revisit RGR's 2016 formal analysis of
capabilities and confused deputies \cite{rajani_garg_rezk2016}.  Their
counterexamples are especially useful because they exhibit several ways in
which possession can be insufficient: the deputy may already possess the
privileged capability while the adversary controls a value, a control decision,
or state used later.  We show that all of these can be represented as causal
paths in the event model; whether such a path is an ordinary PIC authority
segment is a separate premise.  RGR's full provenance then has a precise role: it can witness
some of the data/control dependencies that populate the causal graph, while the
PIC theorem itself is stated over the graph independently of that particular
witness mechanism.

\paragraph{Contributions.}
The paper makes the following claims at the formal level.
\begin{enumerate}[leftmargin=*]
  \item It replaces the privilege tuple as the primitive event by a causal
        occurrence $e\in\mathcal E$, and gives the consistent representation
        $\widehat E\subseteq E\times L$ from which PIC's
        $O\times R\times L$ is obtained by authority-behavior projection.
  \item It generalizes the event behavior from $O\times R$ to
        $\mathcal B$, giving the event-level basis of COSS without introducing
        a second temporal axis.
  \item It extends ordinary PIC authority continuity to causal DAGs, proving
        ancestor-bounded authority, common-root branch separation, and the
        conservative ordinary-join rule
        $C(e)\subseteq\bigcap_{p\in\Pred(e)}C(p)$.
  \item It defines causal/model completeness for protected confused-deputy
        reasoning and proves a universal schema: every causally represented,
        mediated PIC lineage-form confused-deputy effect on an ordinary segment
        is incompatible with ordinary PIC continuity.
  \item It shows that temporal delay, common-root fan-out, and multi-source
        concurrency do not form separate CDA ontologies: they are path length,
        branch identity, and DAG-join structure in the same causal event model.
  \item It rereads RGR's explicit designation, value attack, implicit control
        influence, and initial-heap attack as causal-path instances, while
        preserving their original FPs theorem and terminology.
  \item It gives a conservative event refinement of RGR FPs in which explicit
        and implicit provenance edges can serve as witnesses for causal edges,
        without attributing PoR, PoC, or event-lineage machinery to RGR.
  \item It retains the COSS product construction, Product Continuity,
        projection/factorization results, heterogeneous translations, and
        explicit authorized relaxations as the general multi-coordinate
        extension of PIC.
\end{enumerate}

\section{From PIC Privilege Events to Causal Occurrences}
\label{sec:causal-pic}

\subsection{Lamport-style causal event structure}

Let $\mathcal E$ be the set of execution occurrences in a modeled run and let
$E\subseteq\mathcal E$ be the occurrences retained by the claimed security
model.  The security-relevant causal graph is
\begin{equation}
  \to_c\;\subseteq E\times E,
\end{equation}
an acyclic represented-causality relation.  An edge may correspond to an
immediate machine-level cause or may collapse intervening occurrences omitted
from $E$, but it must be causally sound for the modeled execution.  Write
$\hb=\to_c^{+}$ for its transitive closure.  This is a happened-before-style
\emph{strict causal partial order}: no global physical clock or total order is
assumed \cite{lamport1978}.  Sequential execution is a special case in which
relevant events form a chain; concurrent or distributed execution is generally
a DAG.

For $e\in E$, define its causal history
\begin{equation}
  \Hist(e)=\downarrow e
  :=\{x\in E\mid x=e\ \lor\ x\hb e\}.
  \label{eq:hist-downset}
\end{equation}
Let $L=\{\Hist(e)\mid e\in E\}$ be the set of represented causal histories.
In the original linear PIC model, a represented history is a chain.  In the
DAG generalization used here, $\Hist(e)$ may itself contain a fork and, after a
join, several incomparable ancestors.  Thus $L$ denotes causal histories, not
independent clocks and not merely process-local traces.

A consistent causal cut is a set $K\subseteq E$ that is downward closed:
\begin{equation}
  e\in K\ \land\ x\hb e\Longrightarrow x\in K.
  \label{eq:consistent-cut}
\end{equation}
A cut is the correct mathematical analogue of a coherent execution snapshot.
It may contain tips of several incomparable branches; no wall-clock ordering
between those tips is implied.  Any total serialization chosen for analysis
and required to respect $\hb$ is a linear extension of the causal partial order;
such a serialization observes the execution without adding causal ancestry.
The word \emph{global} below therefore means that $\hb$ ranges over the whole
modeled execution, not that the model assumes a globally synchronized physical
clock.

The consistent event-lineage space is
\begin{equation}
  \widehat E
  :=\{(e,\ell)\in E\times L\mid \ell=\Hist(e)\}.
  \label{eq:ehat}
\end{equation}
Because $\Hist$ is a function of the occurrence, the map
$e\mapsto(e,\Hist(e))$ is a bijection from $E$ to $\widehat E$.  Thus $E\times
L$ is useful as an explicit occurrence/history representation, but the history
is not an independent clock.

PoR is the continuation witness used by PIC.  At the semantic level,
\begin{equation}
  \PoR(e,e')\Longrightarrow e\to_c e'.
  \label{eq:por-sound}
\end{equation}
A concrete system may realize this witness with signatures, attestations,
runtime state, messages, provenance evidence, or another sound mechanism.  The
relation $\to_c$ is therefore the causal edge relation of the represented
security graph; all state maps used below are defined on its vertices $E$.

\subsection{PIC as an authority projection of the causal event space}

Let
\begin{equation}
  \Beh_{\mathrm{auth}}:E\rightharpoonup O\times R
\end{equation}
map an authority-relevant effect occurrence to the operation/resource behavior
it realizes.  The original PIC authority event
\begin{equation}
  (o,r,\ell)\in O\times R\times L
\end{equation}
is recovered from $e\in E$ whenever
$\Beh_{\mathrm{auth}}(e)=(o,r)$ and $\ell=\Hist(e)$.  Formally, PIC's event
space is the image of the partial projection
\begin{equation}
\begin{aligned}
  \Psi_{\mathrm{auth}}:\widehat E &\rightharpoonup O\times R\times L,\\
  \Psi_{\mathrm{auth}}(e,\Hist(e))
   &= (\Beh_{\mathrm{auth}}(e),\Hist(e)).
\end{aligned}
\label{eq:pic-projection}
\end{equation}
Hence $O\times R\times L$ is not a second execution ontology.  It is the
behavior/history view needed by the authority model.

Each occurrence also carries an authority context
\begin{equation}
  C:E\to\powerset(O\times R).
  \label{eq:authority-state-map}
\end{equation}

\begin{definition}[Ordinary event and ordinary edge]
\label{def:ordinary-event}
A non-origin occurrence $e\in E$ is \emph{ordinary} when all of its
security-relevant incoming causes are handled as non-expansive PIC
continuations.  If any incoming cause participates in an authorized grant,
composition, or other non-conservative authority change, then
$e$ is classified as a non-ordinary composition/relaxation event.  We write
\begin{equation}
  \Ord(p,e)
  \quad\Longleftrightarrow\quad
  p\to_c e\ \land\ e\text{ is ordinary}.
  \label{eq:ord-edge}
\end{equation}
Thus ordinariness is resolved at the child occurrence, not independently per
incoming edge; mixed ordinary/non-ordinary joins are classified as
non-ordinary events.
\end{definition}

For an ordinary causal edge, PIC continuity is
\begin{equation}
  \Ord(e,e')
  \Longrightarrow C(e')\subseteq C(e).
  \label{eq:authority-edge}
\end{equation}
This is exactly the original non-expansion rule, now stated directly on causal
occurrences.

\subsection{From authority behavior to general behavior}
\label{sec:leap}

COSS changes the behavior projection, not the causal time.  Fix a finite
family of behavior spaces $(\mathcal B_j)_{j\in J}$ and define the typed product
\begin{equation}
  \mathcal B:=\prod_{j\in J}\mathcal B_j.
\end{equation}
If only some facet combinations are meaningful, let
$\mathcal B^*\subseteq\mathcal B$ denote that admissible behavior universe;
otherwise take $\mathcal B^*=\mathcal B$.  Let
\begin{equation}
  \Beh:E\to\mathcal B^*
\end{equation}
record the behavior realized by an occurrence.  The corresponding explicit
representation is
\begin{equation}
  \widehat E_{\mathcal B}
  =\{(\Beh(e),\Hist(e))\mid e\in E\}
  \subseteq\mathcal B^*\times L.
  \label{eq:behavior-lineage-image}
\end{equation}
With $\mathcal B=O\times R$ this recovers PIC.  With
$\mathcal B=\prod_j\mathcal B_j$, the same occurrence can expose authority,
flow, provenance, identity transition, geography, or other behavior facets.
What occurred and when/why it occurred are therefore separated cleanly:
$\Beh(e)$ is the behavior, while $\Hist(e)$ is its causal history.

\subsection{Paths, common roots, concurrency, and joins}

A lineage \emph{path} is totally ordered by ancestry, but the full execution
need not be.  The first distinction needed for a DAG is between a common root
and a common continuation.  Let
\begin{equation}
  r\hb a,\qquad r\hb b,
  \qquad
  a\not\hb b,\qquad b\not\hb a.
  \label{eq:fork-shape}
\end{equation}
Then $a$ and $b$ share the root $r$ but lie on distinct causal branches.

\begin{proposition}[Common ancestry does not identify causal histories]
\label{prop:common-root}
Under Equation~\eqref{eq:fork-shape},
\begin{equation}
  \Hist(a)\neq\Hist(b),
\end{equation}
even though $r\in\Hist(a)\cap\Hist(b)$.
\end{proposition}

\begin{proof}
By reflexivity of the history definition, $a\in\Hist(a)$.  Since $a$ and $b$
are incomparable, $a\notin\Hist(b)$.  Hence the histories differ.  The common
root belongs to both because $r\hb a$ and $r\hb b$.
\end{proof}

Thus
\begin{equation}
  \boxed{\text{same root}\ \not\Rightarrow\ \text{same causal continuation}.}
  \label{eq:same-root-not-same-lineage}
\end{equation}
Co-residence in one process or one address space is weaker still: physical
availability of two objects to one executor does not merge the causal histories
by which those objects arrived.

Suppose now that two branch tips are both immediate causes of a later event
$e$:
\begin{equation}
  a\to_c e,
  \qquad
  b\to_c e.
  \label{eq:causal-join-shape}
\end{equation}
The event $e$ is a causal join.  The two incoming branches may have a common
root as above, or they may have unrelated roots; the continuity question is the
same at the join.

This matters because authority can attenuate differently after a fork.  For
example, let two privileges $\alpha$ and $\beta$ be available at $r$, while
ordinary continuations specialize the branches so that
\begin{equation}
  C(a)=\{\alpha\},
  \qquad
  C(b)=\{\beta\}.
  \label{eq:branch-specialization}
\end{equation}
Both branches are valid descendants of the same root.  Their later meeting does
not, by itself, authorize the union $\{\alpha,\beta\}$.  The security question
is whether an ordinary join may recreate authority that one of its causal
parents no longer carries.  Section~\ref{sec:universal-cda} answers no.

Operationally, $C(e)$ is a \emph{causal authority envelope}: the effects that
the continuation at occurrence $e$ is permitted to realize.  It is not the
union of capabilities, credentials, references, or other objects physically
reachable in the executor's memory.  This distinction lets PIC reason about a
process that simultaneously stores objects from several causal branches
without treating co-residence as proof that their authority may be combined.

\section{Causal PIC and Universal Confused-Deputy Exclusion}
\label{sec:universal-cda}

\subsection{Ordinary joins cannot silently union authority}

For $e\in E$, let $\Pred(e)=\{p\in E\mid p\to_c e\}$ be its immediate causal
predecessors in the security-relevant graph.

\begin{definition}[Ordinary causal join]
A multi-parent occurrence $e$ is an \emph{ordinary causal join} when $e$ is an
ordinary event in the sense of Definition~\ref{def:ordinary-event}.  Equivalently,
\begin{equation}
  \forall p\in\Pred(e),\quad \Ord(p,e).
\end{equation}
An occurrence with any non-ordinary incoming security-relevant cause is
classified as an explicit composition/relaxation event; Equation~\eqref{eq:authority-edge}
is then not imposed on any of its incoming edges merely by calling another
parent ``ordinary.''
\end{definition}

\begin{proposition}[Conservative ordinary-join authority]
\label{prop:ordinary-join}
If $e$ is an ordinary causal join, then
\begin{equation}
  C(e)\subseteq\bigcap_{p\in\Pred(e)} C(p).
  \label{eq:join-intersection}
\end{equation}
\end{proposition}

\begin{proof}
For every $p\in\Pred(e)$, Equation~\eqref{eq:authority-edge} gives
$C(e)\subseteq C(p)$.  Membership in every parent set is exactly membership in
their intersection.
\end{proof}

The proposition is deliberately conservative.  It does not claim that every
legitimate multi-party composition must use set intersection.  It states what
counts as \emph{ordinary continuation}.  If a system intentionally combines
grants so that the child may exercise authority absent from one parent, that
event is security-relevant precisely because it changes the continuity bound;
it must therefore be represented as an explicit authorized
composition/relaxation rather than hidden inside ordinary continuation.  The
rule applies equally when the parents have unrelated origins and when they are
branches of the same earlier root.

For the common-root specialization of
Equation~\eqref{eq:branch-specialization}, an ordinary join satisfies
\begin{equation}
  C(e)\subseteq C(a)\cap C(b)=\varnothing.
  \label{eq:common-root-empty-join}
\end{equation}
The empty intersection is not a claim that either branch is invalid.  It says
that no authority remains valid as an \emph{ordinary consequence of both
branches simultaneously}.  Any desired recombination must be represented as a
separate authorized composition event.

\begin{corollary}[No authority resurrection at an ordinary join]
\label{cor:no-resurrection}
Let $e$ be an ordinary causal join and let $(o,r)$ be a privilege.  If there is
any $p\in\Pred(e)$ for which $(o,r)\notin C(p)$, then
\begin{equation}
  (o,r)\notin C(e).
\end{equation}
Thus common ancestry cannot restore a privilege that is absent from one of the
causal branches participating in the join.
\end{corollary}

\begin{proof}
By Proposition~\ref{prop:ordinary-join}, $C(e)\subseteq C(p)$ for every
$p\in\Pred(e)$.  The claim follows immediately.
\end{proof}

\begin{definition}[Ordinary authority segment on a DAG]
\label{def:ordinary-segment}
For $q,e\in E$ with $q\hb e$, write $\OrdSeg(q,e)$ when there exists a causal
path
\begin{equation}
  q=e_0\to_c e_1\to_c\cdots\to_c e_n=e
\end{equation}
such that $\Ord(e_{i-1},e_i)$ for every $1\le i\le n$.  Because ordinariness is
resolved at the child occurrence, a path cannot pass through a mixed or
non-conservative join and still satisfy $\OrdSeg$.
\end{definition}

\begin{theorem}[Ancestor-bounded authority on a causal DAG]
\label{thm:ancestor-bound}
Let $q,e\in E$ with $q\hb e$.  If $\OrdSeg(q,e)$, then
\begin{equation}
  C(e)\subseteq C(q).
  \label{eq:req-bound}
\end{equation}
If every causal ancestor $a$ of $e$ also satisfies $\OrdSeg(a,e)$, the same
bound holds simultaneously for every such ancestor.
\end{theorem}

\begin{proof}
Choose a path $q=e_0\to_c e_1\to_c\cdots\to_c e_n=e$.
Equation~\eqref{eq:authority-edge} gives
$C(e_n)\subseteq C(e_{n-1})\subseteq\cdots\subseteq C(e_0)$.  Transitivity of
set inclusion yields $C(e)\subseteq C(q)$.  The simultaneous statement follows
by applying the same argument to each ancestor path; at a multi-parent ordinary
join, Proposition~\ref{prop:ordinary-join} preserves all parent bounds.
\end{proof}

\subsection{Causal completeness and effect mediation}

Let $Req\subseteq E$ be request occurrences and $Eff\subseteq E$ protected
effect occurrences, and let $\mathcal D$ be the set of executor instances.
Use the typed relations and maps
\begin{align}
  \CausedBy &\;\subseteq Eff\times Req,\\
  \Accepted &\;\subseteq Eff,\\
  \Realize &:Eff\to O\times R,\\
  \Exec &:Eff\to\mathcal D,\\
  \IndAuth &\;\subseteq\mathcal D\times(O\times R).
\end{align}
Here $\CausedBy$ is the semantic request/effect causation relation for the
stated threat model, $\Accepted(e)$ abbreviates $e\in\Accepted$, $\Realize$
maps an effect to the operation and actual resource it realizes, and
$\IndAuth(D,p)$ states that executor $D$ independently possesses privilege
$p\in O\times R$.

\begin{definition}[CDA-complete causal event representation]
\label{def:event-complete}
A modeled execution is \emph{CDA-complete} for a stated threat model when:
\begin{enumerate}[label=(C\arabic*),leftmargin=*]
  \item every protected effect in scope is represented by an event in $Eff$;
  \item whenever $\CausedBy(e,q)$ holds for an in-scope request $q$ and
        protected effect $e$, the represented causal graph preserves that
        semantic causation as ancestry:
        \begin{equation}
          \CausedBy(e,q)\Longrightarrow q\hb e.
          \label{eq:causedby-ancestry}
        \end{equation}
\end{enumerate}
The implication in (C2) is intentionally one-way.  Causal ancestry can exist
without the earlier occurrence being a semantic cause of the particular effect
under study; the theory does not identify all happened-before pairs with
request/effect causation.
\end{definition}

When a claim additionally reasons about the provenance of independently held
authority, we use the following optional strengthening.
\begin{definition}[Authority-source completeness]
An execution is \emph{authority-source complete} for a privilege $(o,r)$ at an
effect $e$ when every in-scope independent authority source that contributes
that privilege is represented by an occurrence or initial causal state in the
ancestry relevant to $e$.
\end{definition}

Condition (C2) is a representation-completeness condition, not an additional
security policy.  Full provenance, trusted runtime tracing, message causality,
explicit read-from edges, or another mechanism may supply evidence that the
represented ancestry is faithful; Section~\ref{sec:fp-axis} examines one such
mechanism.

Protected-effect mediation is a separate enforcement premise:
\begin{equation}
\begin{aligned}
  &e\in Eff\land\Accepted(e)\land\Realize(e)=(o,r)\\
  &\hspace{3em}\Longrightarrow (o,r)\in C(e).
\end{aligned}
\label{eq:effect-admission}
\end{equation}

\begin{proposition}[Represented causal attribution]
\label{thm:derived-ca}
In a CDA-complete causal event representation,
\begin{equation}
  \CausedBy(e,q)\Longrightarrow q\hb e
\end{equation}
and therefore $q\in\Hist(e)$.
\end{proposition}

\begin{proof}
The ancestry implication is condition~(C2).  By the definition of $\Hist(e)$
in Equation~\eqref{eq:hist-downset}, $q\hb e$ implies $q\in\Hist(e)$.
\end{proof}

The proposition records the model/semantics interface rather than deriving
causality from an unrelated label.  A concrete implementation must justify
that the represented graph is faithful enough for condition~(C2).

\subsection{Request-bounded effects}

\begin{lemma}[Request-origin authority bound]
\label{lem:req-bound}
If $\CausedBy(e,q)$ and $\OrdSeg(q,e)$, then
\begin{equation}
  C(e)\subseteq C(q).
\end{equation}
\end{lemma}

\begin{proof}
By Proposition~\ref{thm:derived-ca}, $q\hb e$.  The premise $\OrdSeg(q,e)$
then lets us apply Theorem~\ref{thm:ancestor-bound}.
\end{proof}

\begin{theorem}[Request-bounded effect theorem]
\label{thm:req-effect}
Let $q$ be a request occurrence and $e$ an accepted protected effect.  Then
\begin{equation}
\begin{aligned}
  &\CausedBy(e,q)\land\OrdSeg(q,e)\land\Accepted(e)\\
  &\qquad\land\Realize(e)=(o,r)
  \Longrightarrow (o,r)\in C(q).
\end{aligned}
\label{eq:req-effect}
\end{equation}
\end{theorem}

\begin{proof}
Lemma~\ref{lem:req-bound} gives $C(e)\subseteq C(q)$.  Effect admission
\eqref{eq:effect-admission} gives $(o,r)\in C(e)$.  Hence $(o,r)\in C(q)$.
\end{proof}

\subsection{Lineage-form confused deputy and branch-divergent witnesses}

\begin{definition}[Causal lineage-form confused deputy]
\label{def:cda-pic}
For $q\in Req$, $e\in Eff$, $D\in\mathcal D$, and
$p\in O\times R$, define
\begin{equation}
\begin{gathered}
  \CDApic(q,e,D,p)\Longleftrightarrow\\
  \CausedBy(e,q)\land\Accepted(e)\\
  {}\land\Realize(e)=p\land\Exec(e)=D\\
  {}\land p\notin C(q)\land\IndAuth(D,p).
\end{gathered}
\label{eq:cda-def}
\end{equation}
This is the lineage-form authority-mismatch property excluded below on ordinary
request-to-effect segments; it is not identified with RGR's extensional
$\CDAF$ definition.
\end{definition}

When the model also records the causal provenance of the deputy's independent
authority, let
\begin{equation}
  \AuthSource\;\subseteq E\times\mathcal D\times(O\times R)
\end{equation}
and write $\AuthSource(a,D,(o,r))$ when occurrence $a$ contributes $D$'s
independent authority for $(o,r)$.  If an implementation models pre-execution
state as a distinguished initial source, that source is included in $E$.  If $\AuthSource(a,D,(o,r))$ and $a\hb e$, then the same
CDA instance has an explicit branch/source witness.  The source $a$ may share
an older root with $q$ and still belong to a distinct continuation after a
fork.  This structural witness is useful for join analysis but is not required
for the contradiction in Theorem~\ref{thm:total-cda}.

\begin{theorem}[Universal causal-PIC exclusion on ordinary segments]
\label{thm:total-cda}
For every system whose protected execution has a faithful CDA-complete causal
event representation, whose ordinary causal edges satisfy PIC authority
continuity, and whose protected effects satisfy
Equation~\eqref{eq:effect-admission},
\begin{equation}
\begin{aligned}
  \forall q\in Req,\;\forall e\in Eff,\;\forall D\in\mathcal D,\;
  \forall p\in O\times R:\qquad\\
  \CDApic(q,e,D,p)\land\OrdSeg(q,e)\Longrightarrow\bot.
\end{aligned}
\label{eq:cda-ordinary-impossible}
\end{equation}
\end{theorem}

\begin{proof}
Fix arbitrary $q,e,D,p$ satisfying the antecedent.  By
$\CDApic(q,e,D,p)$ we have $\CausedBy(e,q)$, $\Accepted(e)$, and
$\Realize(e)=p$.  Together with $\OrdSeg(q,e)$,
Theorem~\ref{thm:req-effect} yields $p\in C(q)$.  The same CDA predicate
requires $p\notin C(q)$.  Hence $p\in C(q)\land p\notin C(q)$, a
contradiction.  Independent authority held by $D$ cannot enlarge the
request-carried authority through ordinary continuation; at a multi-source
join Proposition~\ref{prop:ordinary-join} prevents silent union.  A legitimate
non-conservative combination must cross an explicit authorized composition or
relaxation boundary and is therefore outside the ordinary-segment antecedent.
\end{proof}

\begin{corollary}[Temporal distance does not create a new CDA class]
Suppose $q\hb e$ and an ordinary request-to-effect path witnessing
$\OrdSeg(q,e)$ contains arbitrarily many intermediate events or arbitrarily long
wall-clock delay.  The conclusion of Theorem~\ref{thm:req-effect} is unchanged.
\end{corollary}

\begin{proof}
The proof depends only on causal ancestry and transitivity of set inclusion,
not on path length or physical time elapsed.
\end{proof}

\begin{corollary}[Temporal proximity does not create causal ancestry]
\label{cor:temporal-proximity}
Let $e_1,e_2\in E$ be incomparable under $\hb$:
\begin{equation}
  e_1\not\hb e_2\qquad\land\qquad e_2\not\hb e_1.
\end{equation}
Wall-clock proximity, simultaneous observation, co-membership in a consistent
cut, or placement next to one another in a total serialization does not create
a causal relation between them.  In particular, such observations do not merge
their causal histories or their authority justifications.
\end{corollary}

\begin{proof}
The relation $\hb$ is the transitive closure of the modeled immediate-causality
relation $\to_c$.  A consistent cut records events without adding edges, and a
total serialization that respects $\hb$ is only a linear extension of the same
partial order.  Neither operation changes $\to_c$ or its transitive closure.
Hence incomparable events remain incomparable, and their histories remain
distinct unless a later modeled causal event joins them.
\end{proof}

The two temporal corollaries are dual: causal ancestry survives arbitrary
physical delay, while physical proximity does not manufacture ancestry.

The main implication is therefore
\begin{equation}
\boxed{
\begin{gathered}
\text{causally complete protected event model}\\
+\ \text{ordinary PIC authority continuity}\\
+\ \text{protected-effect mediation}\\
+\ \text{ordinary request-to-effect segment}\\
\Longrightarrow\\
\text{every represented PIC lineage-form CDA effect}\\
\text{is inadmissible along ordinary continuation.}
\end{gathered}}
\label{eq:universal-cda-chain}
\end{equation}
This is a universal theorem schema over represented systems, not a claim that
an omitted physical cause has somehow been modeled.

\section{Rajani--Garg--Rezk: What Capabilities and Full Provenance Prove}
\label{sec:rgr}

Rajani, Garg, and Rezk (RGR) study access control, capabilities, provenance
tracking, and confused-deputy attacks in a small imperative region calculus
\cite{rajani_garg_rezk2016}.  Their result is important here because it gives a
formal taxonomy of attacks that possession-style capability checks do and do
not prevent.

Their capability semantics implements Miller et al.'s Property~A, ``no
designation without authority,'' by preventing a region from obtaining a write
capability it is not authorized to wield.  RGR show, however, that this does
not imply CDA-freedom in general.  The capability semantics blocks the
classical \emph{explicit designation} case in which the attacker provides the
privileged capability/reference itself, but it does not block all cases in
which the deputy already possesses the privileged capability and the attacker
influences how it is used.

Their Examples~2--4 distinguish three such shapes.  A \emph{value attack}
lets the adversary control a value that the privileged deputy later writes to a
high-integrity reference.  Their Example~3, \emph{Implicit influence}, is a
control-flow attack in which an adversary-controlled condition determines which
privileged write occurs.  An
\emph{initial-heap attack} lets a privileged capability already present in
state be used later under adversarial influence.  RGR call these forms of
\emph{implicit designation}.  Their Theorem~4 obtains CDA-freedom with
endorsement for capability semantics only under the stated $\mathrm{nihrH}$ and
$\mathrm{nihrP}$ restrictions on the initial heap and non-endorsed high regions.

RGR next add provenance tracking.  Their explicit-only semantics (EPs) tracks
principals whose references explicitly influence a computed value, but omits
implicit control-flow influence.  Their full provenance semantics (FPs) adds a
program-counter label stack: on entry to an \texttt{if} or \texttt{while}
body, the current $pc$ is combined with the condition label, and the assignment
rule checks target provenance, value provenance, and the current $pc$ against
the owner of the written reference.  In this way FPs enforces both explicit and
implicit integrity dependencies.

Their Theorem~6 states
\begin{equation}
  \CDAF(E_{\rho_A},H,\to_{\FP})
  \quad\text{holds unconditionally}
  \label{eq:rgr-thm6}
\end{equation}
for their region calculus and FPs semantics.  Their proof says that FPs enforces
a strong form of information-flow integrity (their Definition~8), which in turn
implies CDA-freedom with endorsement through their Theorem~7.  Theorem~7 states
that $\mathrm{NI\text{-}A\text{-}E}$ implies $\mathrm{CDAF\text{-}E}$ for each
of their four semantics, for every initial heap; they also state that the
converse does not hold \cite{rajani_garg_rezk2016}.  ``Unconditionally'' is
relative to their formal calculus: it removes the additional heap/program
restrictions needed by their capability and EP results.

RGR do not define PoR, PoC, a distributed lineage protocol, or a Lamport event
DAG.  Their labels are principal-valued abstractions of influence.  The causal
interpretation below is therefore ours, and it is used only to relate their
attack classes and provenance mechanism to the event semantics already defined.

\subsection{RGR attack classes as causal paths}

Under a causally complete embedding, the RGR examples differ in the concrete
witness of ancestry, not in the PIC theorem that applies afterward:
\begin{align}
  \text{explicit:}\quad
  &q\to_c \text{capability transfer}\to_c e,\\
  \text{value:}\quad
  &q\to_c \text{data/state dependency}\to_c e,\\
  \text{control:}\quad
  &q\to_c \text{branch dependency}\to_c e,\\
  \text{initial heap:}\quad
  &q\to_c \text{read-from dependency}\to_c e.
\end{align}
The paths may contain arbitrarily many additional events.  What matters is
$q\hb e$.

\begin{proposition}[RGR-class closure under causal PIC]
\label{prop:rgr-closure}
Consider any execution of the four RGR attack shapes embedded in a CDA-complete
causal PIC model so that the relevant capability-transfer, value, control, or
state dependency is represented in $\hb$.  If $\OrdSeg(q,e)$, then an accepted privileged effect
$\Realize(e)=(o,r)$ requires $(o,r)\in C(q)$.  Therefore an RGR-style attack in which the request
origin lacks $(o,r)$ is inadmissible.
\end{proposition}

\begin{proof}
Each embedding supplies $q\hb e$ and the proposition assumes $\OrdSeg(q,e)$.  Theorem~\ref{thm:req-effect} then gives the
claim independently of which dependency shape witnessed the ancestry.
\end{proof}

This proposition does not identify RGR's extensional $\CDAF$ definition with
the PIC lineage-form definition, nor does it claim that Theorem~\ref{thm:total-cda}
is a proof of RGR's Theorem~6.  It shows something narrower and structural:
once an attack instance is represented as request-to-effect causal ancestry,
PIC excludes the authority mismatch by the same argument for all four shapes.

The branch view also clarifies a subtle LOW/HIGH case.  Saying that an
attacker-controlled LOW value and a HIGH authority object are both ``in the
same execution'' or even descend from the same old root does not decide the
security question.  If they reach an effect through different causal branches,
the join must preserve the authority bound of every branch participating in
that effect.  Conversely, if the HIGH privilege is already valid in the causal
authority envelope of every relevant parent, authority-PIC alone need not reject
the effect merely because LOW influenced its value or control flow.  RGR's
non-interference property can still reject such an influence.  This is why
\begin{equation}
  \mathrm{NI\text{-}A\text{-}E}\Longrightarrow\CDAF
\end{equation}
does not collapse IFC into PIC: non-interference is a stronger information-flow
property, while PIC constrains causal transport of authority.  COSS can carry
both as distinct coordinates on the same causal event structure.

\section{Full Provenance as Evidence on the COSS Causal Axis}
\label{sec:fp-axis}

The universal PIC theorem is stated over the causal event structure and does
not require RGR full provenance.  FPs is nevertheless useful as a concrete
language-level way to expose some of the edges that a causally complete model
must represent, especially explicit data influence and implicit control
influence.

\subsection{Conservative event refinement of FPs}

Consider an RGR FPs execution
\begin{equation}
  \Theta_0\to_{\FP}\Theta_1\to_{\FP}\cdots\to_{\FP}\Theta_n.
  \label{eq:fp-trace}
\end{equation}
Associate a ghost occurrence $e_i\in E$ with each reduction step.  Define an
influence relation $\Influence$ by two generators:
\begin{enumerate}[label=(F\arabic*),leftmargin=*]
  \item if a value, reference, target designation, or heap value produced or
        obtained at $e_i$ is a semantic input to the computation at $e_j$,
        then $e_i\Influence e_j$;
  \item if execution of $e_j$ is control-dependent on a branch or loop
        condition evaluated at $e_i$, then $e_i\Influence e_j$.
\end{enumerate}
The transitive closure is $\Influence^{+}$.

\begin{proposition}[Conservative erasure]
Erasing the ghost event identifiers and $\Influence$ edges yields exactly the
underlying RGR FPs trace, with the same principal-valued labels, $PC$ evolution,
and accepted or blocked assignments.
\end{proposition}

\begin{proof}
The event identifiers and dependency edges occur in no premise of an RGR FPs
reduction rule.  Erasure therefore commutes with each reduction step; induction
on the trace length gives the result.
\end{proof}

\begin{lemma}[Full-provenance influence is causally ordered]
\label{lem:fp-causal}
If the causal event semantics represents producer/consumer and
condition/control dependencies faithfully, then
\begin{equation}
  e_i\Influence e_j\Longrightarrow e_i\hb e_j,
  \qquad
  \Influence^{+}\subseteq\hb.
  \label{eq:fp-suborder}
\end{equation}
\end{lemma}

\begin{proof}
An explicit FPs dependency means the later step consumes a value/reference or
state whose relevant production/read occurred earlier; semantic dependency
therefore places the producer/read in the causal past of the consumer.  An
implicit FPs dependency means the later command is reachable only through a
branch/loop condition whose label contributes to the active $pc$; the condition
occurrence is therefore in the causal past of the controlled occurrence.
Transitivity gives the closure statement.
\end{proof}

The reverse inclusion need not hold.  Causal ordering may exist without data or
control influence.  Thus FPs can witness an influence suborder of the common
execution; it does not define a second time and is not required to make PIC's
authority theorem true.

\subsection{Evidence role in asynchronous and persistent systems}

The same semantic idea can be realized by other witnesses.  If a later event
consumes state or a message produced by an earlier event, a causally complete
system can expose dependencies such as
\begin{align}
  \mathrm{write}_x &\to_c \mathrm{read}_x,\\
  \mathrm{enqueue}_q &\to_c \mathrm{dequeue}_q,\\
  \mathrm{send}_m &\to_c \mathrm{receive}_m,\\
  \mathrm{setTimer} &\to_c \mathrm{fireTimer},\\
  \mathrm{registerCallback} &\to_c \mathrm{invokeCallback}.
\end{align}
A database, queue, timer, callback, shared object, or message broker therefore
does not constitute a new category of confused deputy merely because the
original synchronous stack has disappeared.  The mathematical requirement is
that the cause remain represented in the ancestry of the effect.

\section{Historical Motivation: Different Security Coordinates}

Bell and LaPadula developed a mathematical model of secure computer states and
rules intended to preserve security under state transitions
\cite{bell_lapadula1973,bell_lapadula1973model}.  Denning subsequently formulated secure information
flow using a lattice of security classes, giving a general mathematical view of
systems that restrict information flow \cite{denning1976}.

Capability systems start from a different primitive. Authority is represented
by protected designators that permit operations on objects. In 1984, Boebert
showed that an \emph{unmodified} capability machine, in the absence of
additional mechanisms, cannot enforce the Bell--LaPadula $*$-property in his
model \cite{boebert1984}. His concrete attack lets a low-clearance program place
a read/write capability for a low object into that low object; a later
high-clearance program may read the low object, obtain the capability, and then
combine read access to high data with write access to the low object.  Boebert's
concluding explanation is that, in the pure capability machine he defines, the
right to exercise access carries the ability to propagate that access.

Miller, Yee, and Shapiro revisited this construction directly in
\emph{Capability Myths Demolished} \cite{miller_yee_shapiro2003}.  They agree
that the particular construction described by Boebert fails to enforce the
$*$-property, but argue that the attack assumes capabilities can be transmitted
wherever data can be transmitted.  In partitioned or type-enforced capability
systems, capability transfer and data transfer are distinguishable operations.
Their two-level construction independently controls those operations and is
presented as a counterexample to the broader claim that capability systems as a
class cannot enforce the $*$-property.  The same paper also stresses that a
static access matrix is not a complete description of a security mechanism:
the rules by which access relationships evolve over time materially affect the
security properties of the resulting system.

Miller and Shapiro develop a complementary argument in \emph{Paradigm
Regained} \cite{miller_shapiro2003}.  They distinguish permission from
authority and explicitly make program behavior on permitted causal pathways
part of authority analysis.  Their object-capability model carries overt
causation by messages over permitted paths.  Their examples use Redell's
Caretaker pattern for revocation, factories for overt confinement, and a data
diode for the $*$-properties.  They also
show how multiple independently motivated abstractions can coexist over the
same graph under ``mutually suspicious composition.''

These papers therefore already contain dynamic, behavioral, and causal
reasoning; \Model does not claim otherwise.  The distinction made here is
representational.  In those analyses, causal pathways and abstraction behavior
are derived from the reference graph and program semantics.  \Model instead
makes causal lineage an explicit axis of the security event and transports an
ordered security state along that axis.  Nor does \Model assume that a
capability implementation can contribute only an authority coordinate: a
capability design may itself mediate information flow, transfer type, request
context, or another security distinction.  The general claim is narrower: if an
enforcement decision is invariant under a distinction that changes the target
security verdict, that decision cannot enforce the target property.  A typed
transfer rule or access abstraction may expose exactly the distinction that a
coarser model omitted.

This paper does \emph{not} claim that Boebert, Bell--LaPadula, Denning, Miller,
or Shapiro formulated their work in this product language.  The product view is
a new interpretation intended to explain why apparently conflicting results
can all be correct under different observable state spaces.

\section{Ordered Security-State Systems}

\begin{definition}[Security-state coordinate]\label{def:coordinate}
A security-state coordinate is a tuple
\begin{equation}
  \mathfrak{S}_j=(X_j,\preceq_j,\mathcal{B}_j,\sem{\cdot}_j),
\end{equation}
where:
\begin{itemize}[leftmargin=*]
  \item $X_j$ is a set of security states;
  \item $\preceq_j$ is a partial order interpreted as ``no more permissive
        than'';
  \item $\mathcal{B}_j$ is a universe of behaviors, transitions, principals,
        flows, or other semantically admitted possibilities; and
  \item $\sem{x}_j\subseteq\mathcal{B}_j$ is the semantic extension admitted
        by state $x\in X_j$.
\end{itemize}
The coordinate is \emph{semantically monotone} when
\begin{equation}
  x'\preceq_j x
  \quad\Longrightarrow\quad
  \sem{x'}_j\subseteq\sem{x}_j.
  \label{eq:semantic-monotonicity}
\end{equation}
\end{definition}

The definition deliberately does not require every security system to be born
as a powerset.  A lattice label, a logical predicate, a capability set, a
provenance constraint, or a history-dependent state can all be used if they
admit a sound refinement order.

\begin{remark}[Coordinates are semantic dimensions]
A coordinate is not required to correspond one-to-one with an implementation
mechanism.  One mechanism may realize several security coordinates, as when a
type-enforced capability runtime distinguishes authority from data transfer and
capability transfer; conversely, several mechanisms may jointly realize one
coordinate.  \Model reasons over the security distinctions exposed to the
admission decision, not over product names or implementation categories.
\end{remark}

\begin{remark}[Refinement is not a complete update semantics]
The order $\preceq_j$ is the condition needed for the continuity theorem, not
a claim that every refinement step is reachable or permitted by the concrete
security system.  A coordinate may impose a stricter local transition relation
$U_j\subseteq X_j\times X_j$ satisfying
\begin{equation}
  U_j(x,x')\Longrightarrow x'\preceq_j x.
\end{equation}
All ordinary-continuation results below remain valid when transitions must
satisfy both $U_j$ and the refinement order.  This leaves room for the richer
update rules that distinguish dynamic security mechanisms even when they share
the same static state representation.
\end{remark}

\begin{definition}[Product security state]
For a finite family $J$ of security-state coordinates, define
\begin{equation}
  X \;=\; \prod_{j\in J} X_j,
  \qquad
  \state=(\coord^{(j)})_{j\in J}.
\end{equation}
The product refinement order is
\begin{equation}
  \state'\preceq\state
  \quad\Longleftrightarrow\quad
  \forall j\in J,\;
  \coord'^{(j)}\preceq_j\coord^{(j)}.
  \label{eq:product-order}
\end{equation}
\end{definition}

For a concrete execution, let
\begin{equation}
  \st:E\to X
\end{equation}
assign exactly one product security state to each execution occurrence, and
write $\state_i:=\st(s_i)$. If $\ell\in L$ is a finite causal history with
endpoint $s_n$, define
\begin{equation}
  \state(\ell):=\st(s_n).
  \label{eq:lineage-endpoint-state}
\end{equation}
Thus the state used by an event is a well-defined function of the endpoint
occurrence of its represented history.

\begin{proposition}[Product order]
If every $(X_j,\preceq_j)$ is a partial order, then
$(X,\preceq)$ defined by \eqref{eq:product-order} is a partial order.
\end{proposition}

\begin{proof}
Reflexivity, antisymmetry, and transitivity hold componentwise.
\end{proof}

\begin{remark}[Coordinates need not be independent]
The Cartesian product is an ambient state space, not an assumption that every
combination of coordinate states is meaningful. If cross-coordinate constraints
permit only states in $Y\subseteq X$, the induced order on $Y$ is still a
partial order and Product Continuity applies unchanged to lineages whose states
remain in $Y$. Likewise, if only certain combinations of behavior facets are
meaningful, one may restrict the event universe to
$\mathcal{B}^{*}\subseteq\prod_j\mathcal{B}_j$ and interpret product admission
on $\mathcal{B}^{*}\cap\prod_j\sem{\coord^{(j)}}_j$.
\end{remark}

\section{Causally Ordered Security-State Continuity}

\begin{definition}[Valid \Model transition]
A transition
\begin{equation}
  (s_i,\state_i)\longrightarrow(s_{i+1},\state_{i+1})
\end{equation}
is \Model-valid when
\begin{equation}
  \PoR(s_i,s_{i+1})
  \quad\land\quad
  \state_{i+1}\preceq\state_i.
  \label{eq:valid-cosst}
\end{equation}
Equivalently, every security coordinate refines or preserves its predecessor
state while the execution advances along a verified causal edge.
\end{definition}

\begin{definition}[Product security-state continuity]
A lineage
\begin{equation}
  \Pi=\langle(s_0,\state_0),\ldots,(s_n,\state_n)\rangle
\end{equation}
has product security-state continuity when every adjacent transition is
\Model-valid.
\end{definition}

\begin{theorem}[Product Continuity]
If $\Pi$ has product security-state continuity, then
\begin{equation}
  \state_n\preceq\state_{n-1}\preceq\cdots\preceq\state_0.
\end{equation}
Hence, for every coordinate $j$,
\begin{equation}
  \coord_n^{(j)}\preceq_j\coord_0^{(j)}.
  \label{eq:component-origin-bound}
\end{equation}
If the coordinate is semantically monotone, then
\begin{equation}
  \sem{\coord_n^{(j)}}_j
  \subseteq
  \sem{\coord_0^{(j)}}_j.
\end{equation}
\end{theorem}

\begin{proof}
Every valid transition gives
$\state_{i+1}\preceq\state_i$.  Transitivity of the product order gives
$\state_n\preceq\state_0$.  Projection to coordinate $j$ gives
\eqref{eq:component-origin-bound}; semantic inclusion follows from
\eqref{eq:semantic-monotonicity}.
\end{proof}

The theorem states the general continuity invariant for lineages containing
only ordinary continuations: causal transport cannot introduce a possibility
forbidden by the origin state of any carried security system.

\begin{theorem}[Pathwise Product Continuity on a causal DAG]
\label{thm:dag-continuity}
Let $q,e\in E$ with $q\hb e$. If every edge of a causal path from $q$ to $e$
is an ordinary \Model transition, then
\begin{equation}
  \st(e)\preceq \st(q).
\end{equation}
If every immediate predecessor edge of a multi-parent ordinary join is required
to satisfy the same componentwise non-expansion rule, the bound holds along
every parent path.
\end{theorem}

\begin{proof}
Choose a causal path
$q=s_0\to_c s_1\to_c\cdots\to_c s_k=e$. Each ordinary edge gives
$\st(s_{i+1})\preceq\st(s_i)$. Transitivity yields
$\st(e)\preceq\st(q)$. The multi-parent statement follows because the same
argument applies independently to every immediate-predecessor path.
\end{proof}

\begin{definition}[Product admission]\label{def:product-admission}
The product state $\state=(\coord^{(j)})_{j\in J}$ admits
$b\in\mathcal{B}^*$ iff $b^{(j)}\in\sem{\coord^{(j)}}_j$ for every $j$.
Write
\begin{equation}
  \sem{\state}:=
  \mathcal B^*\cap\prod_{j\in J}\sem{\coord^{(j)}}_j
  \subseteq\mathcal{B}^*.
\end{equation}
\end{definition}

\begin{proposition}[Joint semantic monotonicity]\label{prop:joint}
If every coordinate is semantically monotone, then
$\state'\preceq\state$ implies $\sem{\state'}\subseteq\sem{\state}$.
Consequently, under product security-state continuity,
$\sem{\state_n}\subseteq\sem{\state_0}$.
\end{proposition}

\begin{proof}
$\state'\preceq\state$ gives $\coord'^{(j)}\preceq_j\coord^{(j)}$ for
every $j$, hence $\sem{\coord'^{(j)}}_j\subseteq\sem{\coord^{(j)}}_j$ by
semantic monotonicity, and intersection with the fixed set $\mathcal B^*$ preserves inclusion. The
second claim follows from Product Continuity.
\end{proof}

A later occurrence therefore admits only behaviors that the origin state
admits in every coordinate jointly, not merely coordinate by coordinate.

\section{Why Refinement Can Be Read as Set Inclusion}

The previous theorem is intentionally stated over partial orders rather than
subsets.  Nevertheless, the subset intuition can be recovered for every
partial order.

\begin{definition}[Principal lower-set representation]
For a poset $(X,\preceq)$ define
\begin{equation}
  \downarrow x
  =\{y\in X\mid y\preceq x\}.
\end{equation}
\end{definition}

\begin{theorem}[Principal down-set embedding]
For every poset $(X,\preceq)$, the standard principal down-set map
\begin{equation}
  \eta:X\to\powerset(X),
  \qquad
  \eta(x)=\downarrow x
\end{equation}
is an order embedding:
\begin{equation}
  x'\preceq x
  \quad\Longleftrightarrow\quad
  \downarrow x'\subseteq\downarrow x.
  \label{eq:downset-embedding}
\end{equation}
\end{theorem}

\begin{proof}
If $x'\preceq x$ and $y\preceq x'$, transitivity gives $y\preceq x$, so
$\downarrow x'\subseteq\downarrow x$.  Conversely,
$\downarrow x'\subseteq\downarrow x$ implies $x'\in\downarrow x$ because
$x'\in\downarrow x'$ by reflexivity.  Hence $x'\preceq x$.
\end{proof}

This standard principal down-set embedding formalizes the subset intuition
behind the present work. A security system need not literally store a set, but
its refinement order can be represented faithfully as inclusion among sets of
states no more permissive than the current one. This inclusion must not be
confused with semantic admission: $\downarrow x\subseteq X$ is a set of
\emph{states}, whereas $\sem{x}\subseteq\mathcal{B}$ is a set of
\emph{behaviors}. Semantic monotonicity connects refinement to behavior
inclusion, but the two constructions are distinct. The familiar PIC rule
$C_{i+1}\subseteq C_i$ is therefore one concrete instance of a more general
ordered refinement pattern.

\section{Projection, Quotients, and Coordinate Blindness}

The second PIC paper interprets lineage-forgetting as a quotient and shows that
lineage-invariant decisions factor through the projection that forgets causal
history \cite{gallo_relational2026}.  The same construction applies to arbitrary
products of security-state coordinates.

Let
\begin{equation}
  X=\prod_{j\in J}X_j
\end{equation}
and let $K\subseteq J$.  Define the coordinate projection
\begin{equation}
  p_K:X\to X_K:=\prod_{j\in K}X_j.
\end{equation}

\begin{definition}[$K$-invariant decision]
A decision rule $A:X\to\{0,1\}$ is $K$-invariant when
\begin{equation}
  p_K(x)=p_K(y)
  \quad\Longrightarrow\quad
  A(x)=A(y).
\end{equation}
\end{definition}

\begin{theorem}[Projection Factorization]
A decision rule $A:X\to\{0,1\}$ is $K$-invariant if and only if there exists a
unique rule
\begin{equation}
  \bar A:p_K(X)\to\{0,1\}
\end{equation}
such that
\begin{equation}
  A=\bar A\circ p_K.
  \label{eq:factorization}
\end{equation}
\end{theorem}

\begin{proof}
If $A=\bar A\circ p_K$, equal projections yield equal decisions.  Conversely,
define $\bar A(z)=A(x)$ for any $x$ with $p_K(x)=z$.  $K$-invariance makes the
definition independent of the representative.  Uniqueness follows because
$p_K$ is surjective onto $p_K(X)$.
\end{proof}

\begin{corollary}[Omitted-coordinate impossibility]
Suppose a target security property $P:X\to\{0,1\}$ distinguishes two states
$x,y\in X$ for which $p_K(x)=p_K(y)$.  No $K$-invariant decision rule can equal
$P$ on all of $X$.
\end{corollary}

\begin{proof}
Every $K$-invariant rule must give the same result on $x$ and $y$, while $P$
requires different results.
\end{proof}

This is the general form of the earlier lineage-forgetting observation: a
security decision cannot enforce a distinction that its observation function
erases.

\begin{corollary}[Behavioral-state blindness]\label{cor:behavioral-blindness}
Let $X=X_{\mathrm{prot}}\times X_{\mathrm{beh}}$ separate a primitive
protection-state coordinate from a coordinate describing security-relevant
behavior or an access abstraction.  If a target property $P:X\to\{0,1\}$
satisfies
\begin{equation}
  P(p,h_1)\ne P(p,h_2)
\end{equation}
for some fixed $p\in X_{\mathrm{prot}}$, then no decision rule that factors
through the projection $(p,h)\mapsto p$ can implement $P$ on all of $X$.
\end{corollary}

\begin{proof}
This is Omitted-coordinate Impossibility with
$K=\{\mathrm{prot}\}$.
\end{proof}

This corollary gives an order-independent statement of the blind spot discussed
by Miller and Shapiro: if two systems have the same protection graph but the
behavior of a security-enforcing abstraction changes the authority or flow
verdict, an analysis that observes only the protection graph is necessarily too
coarse for that distinction \cite{miller_shapiro2003}.  It does not imply that
protection-state analysis is unsound when it succeeds; it states only that such
a projection cannot reproduce a behavior-sensitive property in general.

Using the endpoint-state assignment of \eqref{eq:lineage-endpoint-state}, the
\Model admission rule for an event $(b,\ell)$ is
\begin{equation}
  A_{\Model}(b,\ell)=1
  \quad\Longleftrightarrow\quad
  b\in\sem{\state(\ell)}.
\end{equation}
Let $q_L:\mathcal{B}^*\times L\to\mathcal{B}^*$,
$(b,\ell)\mapsto b$, forget the lineage.

\begin{corollary}[Lineage-forgetting blindness in \Model]
\label{cor:coss-lineage}
Let $b\in\sem{\state(\ell)}\setminus\sem{\state(\ell')}$. No decision rule
that factors through $q_L$ equals $A_{\Model}$ on both $(b,\ell)$ and
$(b,\ell')$.
\end{corollary}

\begin{proof}
$q_L(b,\ell)=q_L(b,\ell')$, so every rule factoring through $q_L$ gives
both events the same verdict, while $A_{\Model}(b,\ell)=1$ and
$A_{\Model}(b,\ell')=0$. This is the same fiber argument as Projection
Factorization, now applied to the projection that omits $L$ from the event
space.
\end{proof}

For the authority coordinate alone this is the lineage-invariance result
of PIC \cite{gallo_poc2026}. In \Model it holds for every carried security
system at once: the same behavior can be admissible in one causal history
and inadmissible in another, in authority, in information flow, in
relational admissibility or in any other coordinate, and a decision that
forgets the history cannot tell the two apart.

\subsection{A three-coordinate example}
Consider a three-step agent lineage
\begin{equation}
  s_0\to s_1\to s_2
\end{equation}
carrying authority, information-flow, and geographic coordinates
\begin{equation}
  \state_i=(C_i,F_i,G_i),
  \qquad
  \state_2\preceq\state_1\preceq\state_0.
\end{equation}
At $s_2$, suppose $C_2$ still permits the operation
$(\mathrm{send},\mathrm{reportAPI})$ and $G_2$ admits the agent's current EU
execution region, but $F_2$ excludes the flow facet $f_{\mathrm{ext}}$
representing transmission of the current data to an external sink. Let $b$ be
the event facet tuple whose authority component is that send operation, whose
flow component is $f_{\mathrm{ext}}$, and whose geographic component is the
current EU region. Authority and geography admit their respective facets, but
IFC does not, so
\begin{equation}
  b\notin\sem{\state_2}
\end{equation}
and the product rejects the event. An authority-only decision would admit the
event even though the complete product state rejects it: the authority
coordinate itself remains valid, while the IFC coordinate does not.

Now consider a different causal history $\ell'$ carrying the same authority
and geographic states but an IFC state $F'$ that admits $f_{\mathrm{ext}}$,
for example after a separately authorized declassification. If $\ell$ denotes
the first history ending at $s_2$, the same event facet tuple $b$ may satisfy
\begin{equation}
  b\in\sem{\state(\ell')}
  \qquad\text{while}\qquad
  b\notin\sem{\state(\ell)}.
\end{equation}
The example exhibits both claims of the theory: all relevant security
coordinates constrain one causal occurrence jointly, and a decision that
forgets the lineage cannot distinguish histories whose current admissibility
differs.

\section{Classical Security Models as Coordinates}

\subsection{Authority and capabilities}

Let $O$ be operations and $R$ resources.  Define
\begin{equation}
  X_{\mathrm{auth}}=\powerset(O\times R),
  \qquad
  C'\preceq_{\mathrm{auth}}C
  \Longleftrightarrow C'\subseteq C.
\end{equation}
This coordinate says which authority may participate in the current causal
execution.  It does not identify the executor that happens to possess other
ambient authority.

With this coordinate alone, \Model-validity reduces to
\begin{equation}
  \PoR(s_i,s_{i+1})
  \quad\land\quad
  C_{i+1}\subseteq C_i,
\end{equation}
which is precisely the abstract continuity rule of PIC
\cite{gallo_poc2026}.

This specialization is compatible with the stronger object-capability results
in the literature.  \emph{Capability Myths Demolished} identifies absence of
ambient authority, inseparability of designation and authority, and
access-controlled delegation channels as important properties of
object-capability systems; it also emphasizes that authority conveyed in the
context of a request helps avoid the classical confused-deputy pattern
\cite{miller_yee_shapiro2003}.  \Model does not treat those protections as
missing.  A capability system may realize several security distinctions at
once; the lineage coordinate matters only for a target property whose verdict
also varies with causal execution after the capability-level distinctions have
been fixed.

\subsection{Information-flow admissibility}

Let $D$ be a domain of information sources and sinks, security classes, or
other flow endpoints.  One possible semantic IFC coordinate is the set of
currently admitted flows
\begin{equation}
  F\subseteq D\times D.
\end{equation}
Define
\begin{equation}
  F'\preceq_{\mathrm{ifc}}F
  \Longleftrightarrow F'\subseteq F.
  \label{eq:ifc-subset}
\end{equation}
A Bell--LaPadula or lattice policy can induce such an admissible-flow relation;
this paper does not replace the internal semantics of those models.  It treats
their effective admissibility state as an independent coordinate whose ordinary
continuation must not become more permissive.

An alternative is to take lattice labels or complete IFC policy states as
$X_{\mathrm{ifc}}$ and define $\preceq_{\mathrm{ifc}}$ directly from their
semantic flow sets.  Equation \eqref{eq:ifc-subset} is therefore an example,
not a claim that all IFC implementations literally store $F$.

The policy state $F_i$ should be distinguished from the dynamic provenance of a
particular run. The former answers which flows are admissible; the latter
answers which earlier occurrences actually influenced a value or control
decision in this execution. RGR's FPs supplies a formal influence abstraction
through value labels and the program-counter stack. Section~\ref{sec:fp-axis}
lifts that influence to event resolution without changing the original FPs
accept/reject behavior.

\subsection{Typed data and capability transfer}

The distinction used by Miller, Yee, and Shapiro to revisit Boebert can itself
be represented as an ordered security coordinate without claiming that their
systems used this notation.  Let $\Lambda$ be a set of security domains and
let
\begin{equation}
  K=\{\mathsf{data},\mathsf{cap}\},
  \qquad
  \mathcal{B}_{\mathrm{tr}}=\Lambda\times\Lambda\times K
\end{equation}
classify a transfer by source domain, destination domain, and whether the
transferred value is ordinary data or a capability.  Define
\begin{align}
  X_{\mathrm{tr}}&=\powerset(\mathcal{B}_{\mathrm{tr}}),\\
  Q'\preceq_{\mathrm{tr}}Q
  &\Longleftrightarrow Q'\subseteq Q,
  \qquad
  \sem{Q}_{\mathrm{tr}}=Q.
\end{align}
The coordinate can therefore admit a data transfer between two levels while
rejecting a capability transfer over the same pair of levels.

This captures the specific distinction made in \emph{Capability Myths
Demolished}: Boebert's attack requires a high-clearance subject to obtain, by
reading a low object, a capability that confers write access to that low object;
the later type-enforced construction distinguishes capability movement from
data movement and blocks the cross-level capability transfer
\cite{miller_yee_shapiro2003}.  The coordinate $Q$ is only an illustrative
semantic encoding of that distinction, not a claim that an object-capability
implementation literally stores a set of typed flows.

\subsection{Relational admissibility}

The relational extension of PIC carries a predicate $\rho_i$ over
successor/predecessor profiles and orders predicates by implication
\cite{gallo_relational2026}.  After quotienting by semantic equivalence, define
\begin{equation}
  X_{\mathrm{rel}}
  =\mathrm{Pred}(\mathcal{T}_J)/{\equiv_{\mathrm{sem}}},
\end{equation}
with
\begin{equation}
  [\rho']\preceq_{\mathrm{rel}}[\rho]
  \quad\Longleftrightarrow\quad
  \rho'\Rightarrow\rho.
\end{equation}
Semantically,
\begin{equation}
  \rho'\Rightarrow\rho
  \quad\Longleftrightarrow\quad
  \sem{\rho'}\subseteq\sem{\rho}.
\end{equation}
Thus the existing PIC product
\begin{equation}
  \Gamma_i=(C_i,[\rho_i]_{\mathrm{sem}})
\end{equation}
is already a two-coordinate instance of \Model.

\subsection{Geographic, provenance, and product constraints}

A geographic restriction can be represented by an admissible region set
$G_i^{\mathrm{allow}}$ ordered by inclusion, while the observed current
location $g_i$ is checked for membership.  A software-product or provenance
coordinate can analogously represent the set of admissible artifact identities,
builders, attestations, or derivation histories.  These coordinates do not need
to share an ontology with authority or IFC; they need only expose a sound
refinement order.

This distinction is useful for invocation.  A product can participate in many
runtime invocations, just as one executor can realize many execution
occurrences.  Product identity is therefore not itself an invocation.  A future
DAG extension can treat invocation as a causal join between an authority
lineage and a product/provenance lineage while preserving the coordinates that
justify each parent.

\section{Re-reading the Capability/IFC Tension}

Consider the two-coordinate state
\begin{equation}
  \state=(C,F)
  \in X_{\mathrm{auth}}\times X_{\mathrm{ifc}}.
\end{equation}
A decision that reads only $C$ factors through the projection
\begin{equation}
  p_C(C,F)=C.
\end{equation}
If two states $(C,F_1)$ and $(C,F_2)$ require different information-flow
verdicts, any $C$-invariant mechanism must nevertheless treat them identically.
This is not a statement that capability systems are intrinsically incapable of
information-flow enforcement.  It is a statement about observational
insufficiency: the decision cannot enforce a distinction represented only in a
coordinate it does not observe.

This gives a compact reinterpretation of the historical sequence without
turning it into a contest between security paradigms.  Boebert's oracle does
observe clearance and classification when it grants a capability; the failure
in his construction occurs in subsequent propagation.  Even an oracle that
correctly creates capabilities cannot, in the unmodified machine he defines,
control their further propagation \cite{boebert1984}.  The distinction here is
therefore between the grant decision and what later executions may acquire and
exercise, not between a labeled and an unlabeled grant.

The later capability responses act directly on that propagation assumption.
Miller and Shapiro's data diode transmits data in one direction and capabilities
in neither \cite{miller_shapiro2003}; Miller, Yee, and Shapiro separately show
how distinguishing capability transfer from data transfer blocks the cross-level
capability movement required by Boebert's attack, while remaining within their
reading of his definition \cite{miller_yee_shapiro2003}.  Under a \Model
encoding, an oracle grant that enlarges the carried authority would be an
origination or explicitly authorized relaxation, whereas acquisition of a new
capability during ordinary continuation cannot silently enlarge the authority
coordinate.  This is a \Model interpretation of the security distinction, not
a claim that Boebert or the later capability papers used this state model.

Miller and Shapiro further show that program behavior and access abstractions can
tighten authority beyond what an arrangement-only analysis of the protection
graph predicts \cite{miller_shapiro2003}.  In \Model terms, these constructions
are examples of mechanisms that make additional distinctions observable or
restrict the semantic state on which the decision operates.  Once those
distinctions are observed, the mechanism no longer belongs to the
projection-blind class covered by the impossibility corollary.  The blindness
claim concerns the information read when an event is admitted; it does not say
that the original capability-granting oracle ignored clearance or
classification.

The same reasoning generalizes.  If
\begin{equation}
  \state=(S_1,S_2,\ldots,S_m),
\end{equation}
a policy that factors only through $S_1$ cannot enforce a property whose
verdict varies with $S_2$ while $S_1$ is fixed.  Security models can therefore
be individually valid and still be jointly necessary.

\section{PIC as the Authority Specialization of COSS}

The relationship between \Model and PIC is a generalization by security-state dimension, while the causal-PIC confused-deputy theorem of Section~\ref{sec:universal-cda} remains the authority specialization.

\paragraph{PIC, first form.}
Take a single coordinate
\begin{equation}
  X=X_{\mathrm{auth}}=\powerset(O\times R).
\end{equation}
Then \Model-validity is exactly PoR plus authority attenuation, and Product
Continuity becomes PIC's origin-binding result.

\paragraph{Relational PIC.}
Take
\begin{equation}
  X=X_{\mathrm{auth}}\times X_{\mathrm{rel}}.
\end{equation}
Then the product order is
\begin{equation}
  (C',\rho')\preceq(C,\rho)
  \quad\Longleftrightarrow\quad
  C'\subseteq C\;\land\;\rho'\Rightarrow\rho,
\end{equation}
which is the product attenuation order already used by the relational theory
\cite{gallo_relational2026}.

\paragraph{General form.}
The natural closure is
\begin{equation}
  X=\prod_{j\in J}X_j,
\end{equation}
where authority, IFC, relational admissibility, provenance, geography,
separation-of-duty memory, or another security system may contribute a
coordinate.  PIC is therefore not renamed or invalidated.  It remains the
authority-continuity specialization from which the more general product
construction was discovered.

\section{COSS in the PIC Series}
\label{sec:series}

\paragraph{Origin of the name.}
\Model abbreviates \emph{Causally Ordered Security-State}.  ``Causally
Ordered'' refers to the single causal partial order shared by all represented
security facts; ``Security-State'' refers to the product of ordered security
coordinates carried on that causal structure.  The name records the
construction's origin as a generalization of PIC: PIC's carried authority state
remains the authority coordinate, while other security-state coordinates are
added without introducing a second temporal or causal axis.

\Model is the third step in a sequence of models that share the same basic
shape: authorization is attached to an execution occurrence in causal history,
PoR relates adjacent occurrences, a carried state is ordered by refinement, and
ordinary continuation may preserve or restrict that state but not silently
expand it.  At the level of this continuity construction, each later paper
contains the earlier state discipline as a specialization; this statement does
not imply that the later construction reproduces every implementation theorem
or deployment assumption of the earlier papers.

\begin{table*}[t]
\centering
\small
\begin{tabular}{@{}p{0.15\linewidth}p{0.255\linewidth}p{0.255\linewidth}p{0.255\linewidth}@{}}
\toprule
 & PoC \cite{gallo_poc2026} & Relational PIC \cite{gallo_relational2026} & \Model \\
\midrule
Event domain & $O\times R\times L$ & execution occurrences indexed by causal histories & $\widehat E_{\mathcal B}\subseteq\mathcal{B}^*\times L$ \\
Carried state & $C\subseteq O\times R$ & $(C,[\rho]_{\mathrm{sem}})$ & $\state\in\prod_j X_j$ \\
Order & $\subseteq$ & $\subseteq$ and $\Rightarrow$, componentwise & $\preceq_j$, componentwise \\
Continuity result & origin-bounded authority & Dual Monotonicity and Combined Closure & Product Continuity \\
Separation result & lineage-invariant policies cannot individuate occurrences & quotient factorization and observation-based indistinguishability & Projection Factorization and lineage-forgetting blindness \\
\bottomrule
\end{tabular}
\caption{The PIC series as a layered continuity construction.}
\label{tab:series}
\end{table*}

The containment statements are exact at the level claimed here.  With
$J=\{\mathrm{auth}\}$, the \Model transition rule is PoR plus
$C_{i+1}\subseteq C_i$, the validity rule of the first PoC model.  With
$J=\{\mathrm{auth},\mathrm{rel}\}$ and relational refinement ordered by
implication, the product order is the relational PIC attenuation order
$(C',\rho')\preceq(C,\rho)$ iff $C'\subseteq C$ and
$\rho'\Rightarrow\rho$ \cite{gallo_relational2026}.

The separation results share the same factorization principle but should not be
identified more strongly than their domains permit.  For the authority-only
event space, take $b=(o,r)$ and two histories $\ell,\ell'$ such that
$(o,r)\in C(\ell)$ and $(o,r)\notin C(\ell')$.  The \Model
lineage-forgetting corollary then gives the same required separation as the
first paper's lineage-invariance result.  The relational paper subsequently
shows that the earlier lineage-forgetting projection is one quotient instance
of a more general factorization construction; its executor-profile quotients and
\Model's coordinate projections are applications of the same mathematical
principle, not literally the same projection map \cite{gallo_relational2026}.

The confused-deputy result is strengthened in the present paper. Hardy's
classical example motivates the problem of a deputy exercising authority for
the wrong request context \cite{hardy1988}. The first PoC paper excludes silent
authority import inside a valid lineage through origin-bounded attenuation, and
the relational paper separates possession from admissible request authority
\cite{gallo_relational2026}. Section~\ref{sec:causal-pic} makes the occurrence
itself primitive; Section~\ref{sec:universal-cda} then lifts PIC to causal DAGs
and proves that every represented request-caused protected effect is bounded by
the request origin on ordinary continuation. RGR full provenance is used later
as one evidence mechanism for causal edges, not as a premise of the authority
impossibility theorem.

\section{Heterogeneous Security-State Spaces}

PIC already permits policy translation between heterogeneous operation/resource
spaces \cite{gallo_poc2026}.  The same idea extends componentwise.

At hop $i$, let coordinate $j$ use state space $X_i^{(j)}$ with order
$\preceq_i^{(j)}$.  Let
\begin{equation}
  T_{i\rightarrow i+1}^{(j)}:
  X_i^{(j)}\to X_{i+1}^{(j)}
\end{equation}
be an order-theoretic policy translation.  A translated transition satisfies
\begin{equation}
  \coord_{i+1}^{(j)}
  \preceq_{i+1}^{(j)}
  T_{i\rightarrow i+1}^{(j)}(\coord_i^{(j)}).
  \label{eq:translated-transition}
\end{equation}

If the translations are monotone and composable, define
\begin{equation}
  T_{0\rightarrow n}^{(j)}
  =T_{n-1\rightarrow n}^{(j)}\circ\cdots\circ T_{0\rightarrow1}^{(j)}.
\end{equation}

\begin{proposition}[Translated continuity]\label{prop:translated}
If every $T^{(j)}_{i\rightarrow i+1}$ is monotone and every transition
satisfies \eqref{eq:translated-transition}, then
\begin{equation}
  \coord_n^{(j)}\preceq_n^{(j)}T_{0\rightarrow n}^{(j)}(\coord_0^{(j)}).
\end{equation}
\end{proposition}

\begin{proof}
By induction on $n$. For $n=0$ the claim is reflexivity. Suppose
$\coord_i^{(j)}\preceq_i^{(j)}T_{0\rightarrow i}^{(j)}(\coord_0^{(j)})$.
Monotonicity of $T^{(j)}_{i\rightarrow i+1}$ gives
$T^{(j)}_{i\rightarrow i+1}(\coord_i^{(j)})\preceq_{i+1}^{(j)}
T_{0\rightarrow i+1}^{(j)}(\coord_0^{(j)})$, and
\eqref{eq:translated-transition} with transitivity gives the claim for
$i+1$.
\end{proof}

Proposition~\ref{prop:translated} is order-theoretic: it requires monotonicity
of the translations and proves a bound in the target orders. A stronger claim
that a translation preserves the intended security meaning across heterogeneous
behavior universes requires a separate semantic relation between those
universes. This paper does not define such a cross-domain semantic soundness
condition; an overly permissive semantic translation is therefore a policy or
deployment error rather than a failure of the order-theoretic continuity
theorem.

\section{Ordinary Continuation versus Authorized Relaxation}
\label{sec:relaxation}

Not every legitimate security transition is monotone.  Declassification may
relax an IFC constraint; a new authority grant may enlarge an authority state;
a policy administrator may broaden an admissible geographic region.

\Model therefore distinguishes two classes of transition.

\begin{definition}[Ordinary continuation]
An ordinary continuation is a PoR-linked transition satisfying
$\state_{i+1}\preceq\state_i$ (or the translated form of
\eqref{eq:translated-transition}).
\end{definition}

\begin{definition}[Authorized relaxation event]
An authorized relaxation event is an explicitly distinguished causal event
whose validity is justified by a separate authority or policy rule and which
may replace one or more coordinates by states not below their predecessors.
\end{definition}

The point is not to prohibit declassification, re-authorization, or policy
change.  It is to prevent those events from being confused with ordinary
continuation.  A security state may become more permissive only at a boundary
whose authority to perform that relaxation is itself explicit and auditable.

This mirrors the distinction already present in PIC between attenuation within
one lineage and the creation of new authority at an origin or explicit future
composition boundary.

Relaxation events are still causal events: they remain causally linked to their
represented predecessor(s), while the ordinary monotonicity obligation is
replaced by the separate authorization/composition rule.

\begin{proposition}[Segment continuity]\label{prop:segment}
Let a PoR-linked lineage contain authorized relaxation events at indices
$1\le r_1<\cdots<r_k$, where the event at $r_m$ is the transition from
$s_{r_m-1}$ to $s_{r_m}$, all other transitions being ordinary
continuations. For every $i$, let $r(i)$ be the largest $r_m\le i$, or $0$
if none exists. Then
\begin{equation}
  \state_i\preceq\state_{r(i)}.
\end{equation}
\end{proposition}

The statement assumes a homogeneous state space along the segment. For segments
containing translated hops, the corresponding bound is the one given by
Proposition~\ref{prop:translated}.

\begin{proof}
If $r(i)=i$, the claim is reflexivity. Otherwise, every transition leaving
$s_{r(i)}$ and ending at or before $s_i$ is an ordinary continuation, because
$r(i)$ is the most recent relaxation-entry index. Product Continuity applied
to the segment from $s_{r(i)}$ through $s_i$ gives
$\state_i\preceq\state_{r(i)}$.
\end{proof}

Origin-boundedness is therefore replaced, in the presence of relaxation,
by boundedness from the most recent explicitly authorized relaxation. A
relaxation cannot be introduced by ordinary continuation; it can only
restart the bound at an explicit, auditable point.

\section{Causal DAGs, Joins, and Multiple Causes}
\label{sec:dag}

The causal event model is a DAG rather than a requirement of one linear
history.  There is one partial order for the modeled execution, while different
lineages are different histories or paths inside that order.  In particular,
a common root does not collapse later branches into one continuation
(Proposition~\ref{prop:common-root}).

The authority coordinate admits a sufficient conservative discipline for ordinary
joins: Proposition~\ref{prop:ordinary-join} requires
\begin{equation}
  C(e)\subseteq\bigcap_{p\in\Pred(e)}C(p).
\end{equation}
This is sufficient for the universal confused-deputy theorem because it
prevents authority available from only one incoming branch from being silently
relabelled as authority of the joined effect.  Corollary~\ref{cor:no-resurrection}
adds the stronger common-root reading: authority lost on one branch cannot be
restored merely because another branch retained it or because both branches
share an ancestor.

For the full COSS product, however, not every coordinate must compose by set
intersection.  A future general join operator
\begin{equation}
  \kappa_j:X_j^{k}\to X_j^{\kappa}
\end{equation}
must state the semantics of legitimate multi-parent composition for coordinate
$j$ and must preserve the provenance of the parent states.  If the result is
more permissive than one or more parents, the join is not ordinary continuation
for that coordinate and must be justified as an explicit authorized
composition/relaxation event.

Thus the present paper supplies a sufficient conservative DAG discipline for
ordinary confused-deputy authority mixing, while leaving the algebra of non-conservative multi-coordinate
composition as a broader COSS research problem.  The distinction is important:
no general product-join theorem is needed to prove that an \emph{ordinary}
authority join cannot silently union independent permissions.

\section{Related Work}

\paragraph{Bell--LaPadula.}
Bell and LaPadula model secure system states together with rules of operation
that preserve security under state transitions \cite{bell_lapadula1973model}.
\Model does not replace those rules or claim that Bell--LaPadula is merely a
set-inclusion policy.  Rather, an effective Bell--LaPadula security state may
serve as one coordinate when its local refinement relation and admitted-event
semantics are specified; \Model adds the explicit causal-event axis and the
product transport rule used when that state must coexist with other security
states.

\paragraph{Denning's lattice model.}
Denning organizes security classes as a lattice whose order is justified by the
semantics of permitted information flow \cite{denning1976}.  This is a direct
example of an ordered security domain, but the IFC semantics remain Denning's:
\Model supplies neither the lattice nor a program analysis for discovering
flows.  It supplies a way to carry such an ordered state together with other
coordinates along one causal execution.

\paragraph{Execution-history access control.}
Abadi and Fournet make run-time rights depend on execution history rather than
only the current stack \cite{abadi_fournet2003}.  In their model, ordinary
execution automatically intersects current rights with the static rights of
code that runs, while an explicit special operation may modify current rights,
at most restoring the static rights of that code and subject to further
constraints.  This is structurally close to the distinction here
between ordinary non-expansive continuation and an explicit relaxation event.
The models nevertheless have different domains: their construction computes a
rights state for code execution history, whereas \Model abstracts over an
arbitrary finite product of security-state coordinates and takes sound PoR as
an explicit premise relating the modeled execution occurrences.

\paragraph{Decentralized information-flow control.}
Myers and Liskov define decentralized labels with owner-specific reader sets,
a restriction order that forms a security-class lattice, and explicit
declassification that is legal only when the process has authority to act for
the relevant owner \cite{myers_liskov1997}.  Their restriction/declassification
distinction is therefore an important prior instance of monotone security
restriction plus explicitly authorized relaxation.  \Model treats such an IFC
state as one possible coordinate; it does not replace the label semantics,
acts-for relation, or static flow analysis of that work.

\paragraph{Rajani--Garg--Rezk full provenance.}
RGR develop what they describe as the first formal characterization of
CDA-freedom in a language-based setting, and later give a formal
characterization of a language fragment in which capabilities can provably
prevent CDAs.  They also show that capability semantics does not prevent all
implicit-designation CDAs and introduce explicit-only and full provenance semantics
\cite{rajani_garg_rezk2016}. FPs augments value provenance with a
program-counter label stack so that branch and loop conditions contribute to
the integrity check on later writes. Their Theorem~6 proves CDA-freedom with
endorsement unconditionally for their region calculus, via information-flow
integrity and their Theorem~7. Read from \Model, FPs is a language-level witness for explicit and implicit
influence edges that may populate the causal structure of Section~\ref{sec:causal-pic}.
\Model does not reinterpret their theorem as a protocol result. The new step is
a conservative event refinement that places FPs-style influence on the same
execution occurrences as PoR; the causal-PIC exclusion theorem itself is stated
over the event ancestry independently of this particular witness.

\paragraph{Unified authorization and dependency calculi.}
Flow-limited authorization integrates authorization and information-flow
reasoning for delegation and trust \cite{arden_liu_myers2015}; the Dependency
Core Calculus gives one calculus in which several dependency analyses,
including information flow, are instances \cite{abadi_dcc1999}; and
operating-system DIFC designs track labels and privileges across processes
\cite{asbestos2005,histar2006,flume2007}. These works unify security reasoning
within a language semantics or an operating-system label model. \Model starts
from a different object: authority bound to execution occurrences in a causal
history that may span services and trust domains, with continuity verifiable
at the receiving boundary. The approaches are complementary: a language-based
or operating-system analysis can supply the influence relation, or a security
coordinate, that \Model transports along PoR.

\paragraph{Capability systems and access abstractions.}
Miller, Yee, and Shapiro show that conclusions about capability systems depend
on the capability model and on distinctions such as data transfer versus
capability transfer, designation with authority, and access-controlled
delegation channels \cite{miller_yee_shapiro2003}.  Miller and Shapiro further
show that security-enforcing program behavior and access abstractions can
restrict authority beyond what a protection-graph-only analysis exposes
\cite{miller_shapiro2003}.  \Model's claim is not that these systems lack
context, flow control, or causal reasoning.  Its contribution is a uniform
event/state construction and a factorization result: whenever a target verdict
depends on a distinction erased by an observation, a decision invariant under
that observation cannot reproduce the target verdict.

\section{Semantic Closure and Realization Assumptions}

Theorem~\ref{thm:total-cda} is architecture-independent but not assumption-free.
Its universality is over systems that faithfully realize the causal model.  A
concrete realization must justify four obligations.
\begin{enumerate}[leftmargin=*]
  \item \emph{Causal completeness.} Every protected request-caused effect and
        every causal dependency relevant to the claimed threat model must be
        represented in the event DAG.  This includes branch identity across
        protocol handoffs and, when in scope, dependencies mediated by memory,
        storage, queues, messages, timers, callbacks, or other state.  A cause
        omitted from the model is not covered by the theorem.
  \item \emph{Sound causal evidence.} PoR, provenance, runtime tracing,
        read-from relations, message metadata, or other evidence used to assert
        an edge must be sound for the semantic causality being claimed.
  \item \emph{Complete protected-effect mediation.} Every protected effect in
        scope must be checked against the authority state of its occurrence, as
        required by Equation~\eqref{eq:effect-admission}.
  \item \emph{Explicit non-conservative transitions.} Endorsement,
        declassification, authority grant, or multi-source composition that
        enlarges a coordinate must be separately authorized and represented as
        such rather than hidden inside ordinary continuation.
\end{enumerate}

These obligations explain the relation between the present paper and the
relational PIC paper.  The earlier theory kept Causal Attribution and Complete
Mediation as deployment assumptions.  Here semantic request/effect causation
remains distinct from happened-before, but CDA-completeness requires every
in-scope causal-attribution fact to be preserved as event ancestry; Complete
Mediation is the protected-effect admission rule.  The proof is therefore
closed inside the formal event model once those representation and enforcement
obligations are satisfied; deployment correctness consists in showing that the
implementation faithfully realizes them.

\section{Limitations and Open Questions}

\paragraph{Physical and covert channels.}
The theorem is total over represented PIC lineage-form CDA instances whose
request/effect pair satisfies the ordinary-segment premise.  A physical influence channel omitted from the causal event semantics is
outside the quantification rather than a counterexample inside it.  Claims
about covert or side channels therefore require instrumentation or a semantics
that actually places the relevant influence in the event graph.

\paragraph{Non-conservative product joins.}
The authority rule for ordinary joins is closed by
Proposition~\ref{prop:ordinary-join}.  General COSS coordinates may require
coordinate-specific join algebras for legitimate non-conservative composition.
Those joins must remain explicit so that they cannot masquerade as ordinary
continuation.

\paragraph{Semantic soundness of heterogeneous translations.}
Translated Continuity proves preservation relative to supplied ordered
translation maps.  Relating states and behaviors across different vocabularies
still requires an application-specific semantic relation and soundness
condition.

\paragraph{Order soundness and coordinate choice.}
A mathematically monotone but semantically incorrect refinement order proves
the wrong property.  Likewise, omitting a coordinate is safe only for target
properties invariant under that omission; the projection theorem makes this
boundary explicit.

\paragraph{Implementation evidence.}
The theory is agnostic about whether causal and security evidence is carried in
tokens, attestations, trusted-runtime state, gateways, proof-carrying objects,
dynamic IFC tracking, or another mechanism.  The obligation is semantic
soundness, not a particular representation.

\section{Discussion}

The conceptual move of the paper is to treat the execution occurrence as the
common object.  PIC originally wrote an authority event as $(o,r,\ell)$.  The
causal formulation makes $e\in\mathcal E$ primitive and obtains
$(o,r,\ell)$ by projection.  COSS then replaces the single authority behavior
with a product behavior $b\in\mathcal B$ and carries heterogeneous ordered
security state on the same occurrence.

This makes several distinctions precise.  First, causal time is not the same as
wall-clock time: arbitrary delay does not break ancestry, and a coherent
snapshot is represented by a consistent cut rather than by assuming a global
physical instant.  Second, concurrency does not require two independent
``times'': incomparable branches live in the same partial order and may meet at
a join.  Third, sharing a root does not make those branches the same lineage.
If authority attenuates differently after a fork, the join cannot reconstruct
the root union by ordinary continuation.  Fourth, a branch-divergent confused deputy can be exposed as an
authority-provenance problem at such a join; the core lineage-form theorem itself
requires only request causation, an authority mismatch, and an ordinary segment.  If one causal branch is a request
whose context lacks $(o,r)$ and another branch carries independent deputy
authority for $(o,r)$, an ordinary PIC join cannot silently produce a child
that exercises $(o,r)$ because the child is bounded by every causal
parent/ancestor.  Intentional authority composition is possible, but it must be
explicit and separately authorized.

RGR's results fit this picture without being rewritten.  Their capability
counterexamples identify concrete ways in which the deputy can already possess
a privileged reference while the adversary controls a value, branch, or stored
state.  Their FPs semantics provides an information-flow mechanism that tracks
these influences inside their language.  Causal PIC abstracts one level lower:
once the request-to-effect dependency is represented in the event DAG, the same
ancestor-bound authority theorem applies regardless of the syntactic carrier of
the influence.

The resulting division of labor is therefore:
\begin{quote}
Causal structure answers \emph{which occurrence caused which later
occurrence}; PIC answers \emph{which authority may continue along that causal
structure}; COSS generalizes the carried state to other security dimensions;
and concrete provenance/attestation mechanisms provide evidence that the
modeled causal and state relations hold in an implementation.
\end{quote}

This is why the confused-deputy result is both strong and carefully scoped.  It
is not merely detection: a protected PIC lineage-form CDA instance on an
ordinary request-to-effect segment is not an admissible ordinary PIC state.
But the theorem is only as physically comprehensive as the causal model and
mediation path that a deployment actually realizes.

\section{Conclusion}

This paper starts from the PIC and relational PIC models and makes their causal
object explicit.  A security-relevant machine execution is represented by one
Lamport-style strict causal partial order $(E,\hb)$, with
$E\subseteq\mathcal E$.  Distinct lineages are
distinct causal histories inside that order; they may share a root and still
diverge into incomparable branches.  The consistent pair
$(e,\Hist(e))\in E\times L$ is the canonical occurrence/history object; PIC's
$(o,r,\ell)\in O\times R\times L$ is its authority-behavior projection, and
COSS's $(b,\ell)\in\mathcal B^*\times L$ is the corresponding general
behavior projection.  There is one causal execution, not separate PIC and IFC times.
Here ``one causal execution'' means one causal partial order: distinct lineages
are different histories within the same DAG, not positions on a shared global
timeline.

On that structure ordinary PIC continuity is local and simple:
$C(e')\subseteq C(e)$ on each ordinary causal edge.  Its transitive consequence
is pathwise: whenever $\OrdSeg(q,e)$ holds, $C(e)\subseteq C(q)$.  At a
multi-source ordinary join,
\begin{equation}
  C(e)\subseteq\bigcap_{p\in\Pred(e)}C(p),
\end{equation}
so authority that belongs only to one incoming branch cannot be silently
imported into the joined effect.  In particular,
\begin{equation}
  \text{same root}\not\Rightarrow\text{same continuation},
\end{equation}
and common ancestry cannot resurrect authority attenuated away on a branch.

With protected-effect mediation, this yields
\begin{equation}
\begin{aligned}
  &\CausedBy(e,q)\land\OrdSeg(q,e)\land\Realize(e)=(o,r)\\
  &\qquad\land\Accepted(e)
  \Longrightarrow (o,r)\in C(q).
\end{aligned}
\end{equation}
A lineage-form confused deputy requires the same effect to satisfy
$(o,r)\notin C(q)$.  Therefore the two conditions imply
\begin{equation}
  (o,r)\in C(q)\land(o,r)\notin C(q),
\end{equation}
and hence contradiction.

RGR's capability counterexamples do not create exceptions to this causal
statement.  Explicit designation, value influence, implicit control influence,
and initial-heap influence are different concrete witnesses of request-to-effect
ancestry.  Their full provenance semantics is a powerful language-level
mechanism for exposing explicit and implicit influence and independently proves
CDA-freedom in their region calculus.  In the present theory it can also serve
as evidence for causal edges, but it is not required as the logical foundation
of PIC authority continuity.  Nor does authority-PIC subsume all of
non-interference: if every relevant causal parent already authorizes a HIGH
effect, a LOW-to-HIGH influence can remain an IFC violation without being an
authority-expansion violation.  The two properties are distinct coordinates on
the same causal execution.

The final result is therefore:
\begin{equation}
\boxed{
\begin{gathered}
\text{Every represented PIC lineage-form CDA instance}\\
\text{on an ordinary request/effect segment is}\\
\text{incompatible with ordinary PIC continuity.}
\end{gathered}}
\end{equation}
Equivalently, for every system satisfying causal completeness, protected-effect
mediation, and ordinary PIC authority continuity,
\begin{equation}
\boxed{
\begin{gathered}
\forall q\in Req,\;\forall e\in Eff,\;\forall D\in\mathcal D,\;
\forall p\in O\times R:\\
\CDApic(q,e,D,p)\land\OrdSeg(q,e)\Longrightarrow\bot.
\end{gathered}}
\end{equation}
The quantifier is universal over systems satisfying that representation and
enforcement contract.  The theorem does not assert coverage of physical causes
omitted from the event model, nor does it exclude executions that cross an
explicitly authorized non-conservative authority transition.  It states that
once the relevant causes and protected effects are inside the causal semantics,
no PIC lineage-form confused-deputy instance is an admissible ordinary PIC
continuation.


\begin{thebibliography}{99}

\bibitem{bell_lapadula1973}
D.~E. Bell and L.~J. LaPadula,
``Secure Computer Systems: Mathematical Foundations,''
The MITRE Corporation, ESD-TR-73-278, Vol.~I, 1973.

\bibitem{bell_lapadula1973model}
L.~J. LaPadula and D.~E. Bell,
``Secure Computer Systems: A Mathematical Model,''
The MITRE Corporation, ESD-TR-73-278, Vol.~II, 1973.

\bibitem{denning1976}
D.~E. Denning,
``A Lattice Model of Secure Information Flow,''
\emph{Communications of the ACM}, vol.~19, no.~5, pp.~236--243, 1976,
doi: 10.1145/360051.360056.

\bibitem{boebert1984}
W.~E. Boebert,
``On the Inability of an Unmodified Capability Machine to Enforce the *-Property,''
in \emph{Proceedings of the 7th DoD/NBS Computer Security Conference},
Gaithersburg, MD, USA, Sep.~24--26, 1984, pp.~291--293.
[Online]. Available: \url{https://csrc.nist.gov/files/pubs/conference/1984/09/24/7th-dodnbs-computer-security-conference/final/docs/1984-7th-conference-proceedings.pdf}.

\bibitem{miller_yee_shapiro2003}
M.~S. Miller, K.-P. Yee, and J.~S. Shapiro,
``Capability Myths Demolished,''
Technical Report SRL2003-02, Systems Research Laboratory,
Department of Computer Science, Johns Hopkins University, Mar. 2003.

\bibitem{miller_shapiro2003}
M.~S. Miller and J.~S. Shapiro,
``Paradigm Regained: Abstraction Mechanisms for Access Control,''
in \emph{Advances in Computing Science -- ASIAN 2003: Programming Languages and Distributed Computation},
Lecture Notes in Computer Science, vol.~2896, Springer, 2003,
pp.~224--242, doi: 10.1007/978-3-540-40965-6\_15.

\bibitem{abadi_fournet2003}
M.~Abadi and C.~Fournet,
``Access Control Based on Execution History,''
in \emph{Proceedings of the 10th Annual Network and Distributed System Security Symposium (NDSS)},
2003.

\bibitem{myers_liskov1997}
A.~C. Myers and B.~Liskov,
``A Decentralized Model for Information Flow Control,''
in \emph{Proceedings of the 16th ACM Symposium on Operating Systems Principles (SOSP)},
pp.~129--142, 1997, doi: 10.1145/268998.266669.

\bibitem{rajani_garg_rezk2016}
V.~Rajani, D.~Garg, and T.~Rezk,
``On Access Control, Capabilities, Their Equivalence, and Confused Deputy Attacks,''
in \emph{2016 IEEE 29th Computer Security Foundations Symposium (CSF)},
Lisbon, Portugal, 2016, doi: 10.1109/CSF.2016.18.

\bibitem{arden_liu_myers2015}
O.~Arden, J.~Liu, and A.~C. Myers,
``Flow-Limited Authorization,''
in \emph{2015 IEEE 28th Computer Security Foundations Symposium (CSF)}, 2015.

\bibitem{abadi_dcc1999}
M.~Abadi, A.~Banerjee, N.~Heintze, and J.~G. Riecke,
``A Core Calculus of Dependency,''
in \emph{Proceedings of the 26th ACM SIGPLAN-SIGACT Symposium on Principles of
Programming Languages (POPL)}, 1999.

\bibitem{asbestos2005}
P.~Efstathopoulos, M.~Krohn, S.~VanDeBogart, C.~Frey, D.~Ziegler, E.~Kohler,
D.~Mazi\`eres, F.~Kaashoek, and R.~Morris,
``Labels and Event Processes in the Asbestos Operating System,''
in \emph{Proceedings of the 20th ACM Symposium on Operating Systems Principles
(SOSP)}, 2005.

\bibitem{histar2006}
N.~Zeldovich, S.~Boyd-Wickizer, E.~Kohler, and D.~Mazi\`eres,
``Making Information Flow Explicit in HiStar,''
in \emph{Proceedings of the 7th USENIX Symposium on Operating Systems Design and
Implementation (OSDI)}, 2006.

\bibitem{flume2007}
M.~Krohn, A.~Yip, M.~Brodsky, N.~Cliffer, M.~F. Kaashoek, E.~Kohler, and
R.~Morris,
``Information Flow Control for Standard OS Abstractions,''
in \emph{Proceedings of the 21st ACM Symposium on Operating Systems Principles
(SOSP)}, 2007.

\bibitem{lamport1978}
L.~Lamport,
``Time, Clocks, and the Ordering of Events in a Distributed System,''
\emph{Communications of the ACM}, vol.~21, no.~7, pp.~558--565, 1978,
doi: 10.1145/359545.359563.

\bibitem{hardy1988}
N.~Hardy,
``The Confused Deputy (or Why Capabilities Might Have Been Invented),''
\emph{ACM SIGOPS Operating Systems Review}, vol.~22, no.~4, pp.~36--38, 1988.

\bibitem{gallo_poc2026}
N.~Gallo,
``Proof-of-Continuity: A Temporal Model for Authority Propagation in Distributed Systems and AI Agents,''
arXiv:2607.08906, 2026.

\bibitem{gallo_relational2026}
N.~Gallo,
``Beyond Proof of Possession: A Relational Theory of Authority,''
manuscript, 2026.

\end{thebibliography}
\end{document}